# Deep Learning for Reliable Classification of COVID-19, MERS, and SARS from Chest X-Ray Images

Anas Tahir[a], Yazan Qiblawey[a], Amith Khandakar[a], Tawsifur Rahman[a], Uzair Khurshid[a], Farayi Musharavati[b], M. T. Islam[c], Serkan Kiranyaz[a], Sumaya Al-Maadeed[d], Muhammad E. H. Chowdhury[a]*

[a] Department of Electrical Engineering, Qatar University, Doha-2713, Qatar

[b] Mechanical & Industrial Engineering Department, Qatar University, Doha-2713, Qatar

[c] Department of Electrical, Electronic & Systems Engineering, Universiti Kebangsaan Malaysia, Bangi, Selangor 43600, Malaysia

[d] Department of Computer Science and Engineering, Qatar University, Doha-2713, Qatar

*****Correspondence: Muhammad E. H. Chowdhury; mchowdhury@qu.edu.qa, Tel.: +974-31010775

## CRediT authorship contribution statement

**Anas Tahir:** Methodology, Software, Validation, Formal analysis, Writing - Review & Editing. **Yazan Qiblawey:** Data Curation, Investigation, Resources, Writing - Original Draft, Writing - Review & Editing. **Amith Khandakar:** Methodology, Visualization, Resources. **Tawsifur Rahman:** Methodology, Software, Data Curation, Writing - Original Draft. **Uzair Khurshid:** Visualization, Resources, Writing - Original Draft. **Farayi Musharavati:** Writing - Review & Editing, Supervision, **M. T. Islam:** Writing - Review & Editing, Supervision, **Serkan Kiranyaz:** Writing - Review & Editing, Supervision, Conceptualization, **Sumaya Al Maadeed:** Writing - Review & Editing, **Muhammad E. H. Chowdhury:** Conceptualization, Writing - Review & Editing, Supervision, Project administration.

## Abstract

*Background:* Novel Coronavirus disease (COVID-19) is an extremely contagious and quickly spreading Coronavirus infestation. Severe Acute Respiratory Syndrome (SARS) and Middle East Respiratory Syndrome (MERS), which outbreak in 2002 and 2011, and the current COVID-19 pandemic are all from the same family of coronavirus. This work aims to classify COVID-19, SARS, and MERS chest X-ray (CXR) images using deep Convolutional Neural Networks (CNNs). To the best of our knowledge, this classification scheme has never been investigated in the literature.

*Methods:* A unique database was created, so-called QU-COVID-family, consisting of 423 COVID-19, 144 MERS, and 134 SARS CXR images. Besides, a robust COVID-19 recognition system was proposed to identify lung regions using a CNN segmentation model (U-Net), and then classify the segmented lung images as COVID-19, MERS, or SARS using a pre-trained CNN classifier. Furthermore, the Score-CAM visualization method was utilized to visualize classification output and understand the reasoning behind the decision of deep CNNs.

*Results:* Several Deep Learning classifiers were trained and tested; four outperforming algorithms were reported: SqueezeNet, ResNet18, InceptionV3, and DenseNet201. Original and preprocessed images were used individually and all together as the input(s) to the networks. Two recognition schemes were considered: plain CXR classification and segmented CXR classification. For plain CXRs, it was observed that InceptionV3 outperforms other networks with a 3-channel scheme and achieves sensitivities of 99.5%, 93.1%, and 97% for classifying COVID-19, MERS, and SARS images, respectively. In contrast, for segmented CXRs, InceptionV3 outperformed using the original CXR dataset and achieved sensitivities of 96.94%, 79.68%, and 90.26% for classifying COVID-19, MERS, and SARS images, respectively. The classification performance degrades with segmented CXRs compared to plain CXRs. However, the results are more reliable as the network learns from the main region of interest, avoiding irrelevant non-lung areas (heart, bones, or text), which was confirmed by the Score-CAM visualization.

*Conclusions:* All networks showed high COVID-19 detection sensitivity (>96%) with the segmented lung images. This indicates the unique radiographic signature of COVID-19 cases in the eyes of AI, which is often a challenging task for medical doctors.

# I. Introduction

The World has experienced outbreaks of coronavirus infections during the last two decades: (i) the Severe Acute Respiratory Syndrome (SARS)-CoV outbreak in 2002-2003 from Guangdong, China; (ii) the Middle East Respiratory Syndrome (MERS)-CoV outbreak in 2011 from Jeddah, Saudi Arabia; and (iii) Coronavirus Disease 2019 (COVID-19) or SARS-CoV-2 outbreak from Wuhan, China in December 2019. SARS had quickly spread to 26 countries before being contained after about four months. More than 8,000 people got infected by SARS and 774 died (10%). Since 2004, there have been no reported SARS cases. MERS is a viral respiratory disease that was first reported in Saudi Arabia in September 2012 and has since spread to 27 countries. According to the World Health Organization (WHO) up to date, there are 2,519 confirmed MERS cases and 34% (866) of these patients have died. Even though all three diseases are from the same family of coronavirus [1], the genomic sequence of COVID-19 showed similar but distinct genome composition from its predecessors SARS and MERS [1, 2]. Despite a lower fatality rate of COVID-19, i.e., around 3% [3] when compared to SARS (10%) and MERS (34%), COVID-19 has resulted in many fold deaths (>2.5 million already) than combined deaths of MERS and SARS (around 1700) [4]. The recent outbreak of COVID-19 was and still is an extremely infectious disease that has spread all over the World, forcing the WHO to declare it as a pandemic on 11th March 2020 [5].

These viruses are originated from animals. However, they can be transmitted to humans, which can cause severe and often fatal respiratory disease in their new host. The coronaviruses have a genetic structure that allows them to quickly replicate their presence and weaken the host's antiviral defense mechanisms. The first symptom of SARS-CoV disease is generally high fever (>38°C), and may also be accompanied by headache, malaise, and muscle pain. At the onset of illness, some cases develop mild respiratory symptoms [6]. Typically, rash and gastrointestinal findings are absent, although some patients might have diarrhea during the early feverish stage. Besides, a lower respiratory phase illness begins with the onset of a dry non-productive cough or dyspnea (shortness of breath) with hypoxemia (low blood oxygen levels). The respiratory illness can be severe enough to require intubation and mechanical ventilation. On the other hand, the clinical spectrum of MERS-CoV infection ranges from no symptoms (asymptomatic), mild respiratory symptoms, or acute respiratory distress syndrome (ARDS). In addition, it can develop to complete respiratory failure or death. Typical symptoms of MERS-CoV disease are fever, cough, and shortness of breath. Pneumonia is a common finding, but not always present. Gastrointestinal symptoms, including diarrhea, have also been reported. Severe illness can cause respiratory failure that requires mechanical ventilation and support in an intensive care unit [6]. In general, the symptoms of MERS-CoV disease and COVID-19 are mostly similar, where people with a compromised immune system, elderly people, or people with chronic diseases are at high risk. Besides, severe cases can undergo several organs failure, particularly heart, kidneys, or septic shocks.

Reliable and early detection of COVID-19 has the utmost importance to stop the spread of the virus. Currently, reverse transcription of polymerase chain reaction (RT-PCR) arrays and chest imaging techniques, such as Computed Tomography (CT) scans and chest X-ray (CXR) imaging, are the main diagnostic tools to detect COVID-19. RT- PCR has become the gold standard tool to diagnose COVID 19 [7]. However, RT-PCR arrays have a high false alarm rate caused by sample contamination, damage to the sample, or virus mutations in the COVID-19 genome [8, 9]. Thus, CT scans and CXRs are recommended as an alternative or secondary diagnostic strategy [10]. In fact, several studies suggest performing CT as a secondary test if the suspected patients with shortness of breath or other respiratory symptoms showed negative RT-PCR findings [10, 11]. However, CT is not very convenient due to the contagious nature of the disease, and it is less available in low-resource countries. On the other hand, CXR is widely used as an assistive tool to diagnosis COVID 19 [12]. X-ray imaging is a cheaper, faster, and easily accessible method with a reduced risk of radiation exposure compared to CT [13].

SARS and MERS emerged in 2002 and 2011 long before the recent advancement in Artificial Intelligence (AI) and Deep Learning methods. Therefore, there have been some modest approaches in the literature to provide an automated detection system for SARS or MERS using CXR images. Freedman *et al.* [15] proposed a computer-aided detection system to detect SARS from the infected lung regions from 59 digital X-ray images of SARS patients. Xie *et al.* [14] proposed a computer-aided SARS detection (CAD-SARS) system to discriminate SARS from other types of pneumonia. First, the lung regions were segmented using a multi-resolution active shape model (MRASM), divided into 18 blocks; then, several image features were extracted. Secondly, the classification and regression tree (CART) was used for SARS recognition. The results showed an accuracy of 70.94% for detecting SARS pneumonia. On the other hand, recently, several state-of-the-art deep Convolutional Neural Networks (CNNs) have been reported to detect COVID-19 from the CXR images [13, 15-27]. Sethy *et al.* [15] proposed a cascaded system, where a deep CNN was utilized for feature extraction followed by a support vector machine (SVM) for classification. Eleven pre-trained networks were investigated, where ResNet50 achieved the best performance with 97.2% COVID-19 sensitivity. However, the work was done on a small dataset of 25 Normal and 25 COVID-19 images. Narin *et al.* [13] used a dataset of 50 COVID-19 and 50 Normal images to train and validate ResNet50, InceptionV3, and Inception-ResNetV2 CNN models. ResNet50 model showed the highest classification performance with 98% accuracy and 96% sensitivity for COVID-19 detection. In [16], the authors presented a modified version of DarkNet, to classify a binary class problem (COVID-19 vs Normal) and a multi-class classification problem (COVID-19 vs. non-COVID pneumonia vs. Normal) using 114 COVID-19 CXR images. The sensitivity of COVID-19 detection for 2-class and 3-class problems was 90.65% and 85.35%, respectively. Apostolopoulos *et al.* [17] have used a dataset of 224 COVID-19, 714 Viral or Bacterial pneumonia, and 504 Normal CXR images to train and evaluate MobileNetV2 CNN model. They have achieved 96.7% accuracy and 98.7% sensitivity for detecting COVID-19.

Wang *et al.* [18] proposed COVID-Net, the network was evaluated on a dataset of 5,941 CXR images from 1,203 Normal cases, 931 patients with Bacterial pneumonia, 660 patients with Viral pneumonia, and 45 patients with COVID-19. The COVID-Net achieved 98.9% positive predictive value (PPV) and 91% COIVD-19 sensitivity. Waheed *et al.* [19] proposed a synthetic data augmentation technique to alleviate the scarcity of public data available for COVID-19 X-rays. Auxiliary Classifier Generative Adversarial Network (ACGAN) model was introduced and evaluated over 403 COVID-19, and 721 Normal CXR images. ACGAN model alone yielded 85% accuracy and 69% COVID-19 detection sensitivity while adding the synthetic X-rays boosted the performance to 95% and 90%, respectively. Chowdhury *et al.* [20] created a dataset of 3,487 CXR images, including COVID-19, Viral pneumonia, and Normal cases. The compiled dataset was used to train and validate SqueezeNet, ResNet18, ResNet101, MobileNetV2, DenseNet201, and CheXNet networks for 2-class (COVID vs. Normal) and 3-class (COVID-19 vs. Viral pneumonia vs. Normal) recognition schemes. DenseNet201 showed the best classification performance with 99.7% and 97.9% COVID-19 detection sensitivities for 2-class and 3-class problems, respectively. A compact CNN architecture, called COIVD capsule network (COVID-CAPS), with a low number of trainable parameters was proposed in [21]. The authors used over 94 thousand CXR images to pre-train the proposed framework for classifying five lung abnormalities. The pre-training was followed by training on the target dataset with 358 COVID-19, 8,066 non-COVID pneumonia, and 5,538 Normal CXR images. It was reported that COVID-CAPS achieved accuracy and area under the curve (AUC) of 95.7%/98.3% and 97%/99%, respectively, for without/with pre-training. Apostolopoulos *et al.* [22] trained MobileNetV2 on a 7-class dataset that includes 358 COVID-19, 1,342 Normal, and 1,199 CXR images for five common lung abnormalities. The reported accuracy for the 7-class problem was 87.66%, while for 2-class (COVID vs. other classes), it was boosted to 99.18% with 97.36% COVID-19 detection sensitivity. Another hybrid method has been proposed by Wang *et al.* [28], which used Wavelet Renyi Entropy with Feed-Forward Neural Network and Three-Segment

Biogeography-Based Optimization (3SBBO) algorithm and have reported a mean of accuracy, sensitivity, and F1-Score of 86.12%, 86.40%, and 86.16%, respectively. Wang *et al.*[29] proposed a new CNN model, FGCNet, to learn individual image-level representations. Deep Feature Fusion (DFF) was proposed to fuse individual image-level features and relation-aware features from both Graph Convolutional Network (GCN) and CNN, respectively. FGCNet provides a sensitivity of 97.71% and accuracy of 97.14 %, which outperform other state-of-the-art approaches.

Worldwide researchers have presented numerous clinical and experimental results about SARS and MERS, which could be useful in the fight against COVID-19 [29] [30]. Several studies are available in the literature on investigating the similarities between the genome structures of SARS, MERS, and COVID-19 viruses [8]. However, to the best of our knowledge, COVID-19 distinction from the MERS and SARS, using CXR images has never been investigated. This is a major novelty of this research apart from several other investigations, including image enhancement, Deep Learning algorithms, lung segmentation, and activation mapping visualization to confirm the findings of this experiment. Due to the overlapping pattern of lung infections, it is very challenging for the medical doctors (MDs) to diagnose the lung pneumonia type (COVID-19, MERS and SARS) without the aid of clinical data. SARS epidemic was started in July 2003. However, no new cases have been reported since May 2004 [31]. On the other hand, MERS still exists, as laboratory-confirmed cases were reported by Riyadh in March 2020 [32]. Therefore, investigating the similarities and uniqueness of coronavirus family-members in the eyes of AI can bridge the knowledge gaps and provide MDs with meaningful insights that can help in the diagnosis process.

In this study, we have compiled the largest coronavirus family CXR dataset, so-called QU-COVID-family with 701 CXR images for COVID-19, SARS, and MERS. Then we have investigated several state-of-the-art deep CNNs models for distinguishing COVID-19 from predecessor family members, SARS and MERS. For this purpose, we proposed a robust COVID-19 recognition system using a cascade of two CNN models. The first CNN model identifies lung regions using the U-Net segmentation model, one of the top-performing networks for biomedical image segmentation. Segmented lung CXR images were fine-tuned by the medical doctors and will be released with the QU-COVID-family database. This is the first of its kind ground-truth masks available for COVID-family X-ray images. Next, a pre-trained CNN model classifies the target region as SARS, MERS, or COVID-19. In this study, several deep CNN classifiers (SqueezeNet, ResNet18, InceptionV3, and DenseNet201) were deployed along with different image enhancement techniques in order to determine the best model and configuration for the target problem. Furthermore, the Score-CAM visualization technique was used to confirm that CNN learns reliably from the main region of interest, i.e., segmented lungs, and to identify the potential distinguishable deep-layer features in the CXR images for COVID-19, MERS, and SARS patients.

The rest of the paper is organized as follows: Section II describes the methodology adopted, the experimental setup, and the evaluation metrics used in this study. Section III presents the results and performs an extensive set of comparative evaluations among the employed networks. Accordingly, we discuss and analyze the results, whereas the conclusions are drawn in Section IV.

## II. Methodology

This section is organized as follows: Section A describes the process of creating the QU-COIVD-family benchmark dataset using CXR images for SARS, MERS, and COVID-19 patients. Next, the proposed two-stage COVID recognition system is introduced (Figure 1). Section B presents the pre-processing techniques applied to the CXR images before feeding them to the deep networks, Section C briefly describes the U-Net segmentation model for lung segmentation, and Section D describes four deep classification CNNs (SqueezeNet, ResNet18, InceptionV3, and DenseNet201) used to discriminate between different coronavirus family members. Additionally, in Section E, the Score-CAM visualization method is deployed to provide an interpretable result and investigate the reasoning behind the specific decisions of the deep classification CNNs. Finally, Section F describes the experimental setup of this study.

### *A.* ***Database Description***

Worldwide, the number of infected cases for COVID-19 already exceeds 115 million, and the death toll is around 2.6 million [33]. However, little effort has been made by highly infected countries on sharing clinical and radiography data publicly. Sharing COVID-19 data will help researchers, doctors, and engineers around the world to come up with innovative solutions for early detection. Therefore, we have created a COVID family dataset (QU-COVID-family) for COVID-19, MERS, and SARS with 423, 144, and 133 CXR images, respectively. The dataset was created by utilizing the CXR images available publicly in the published or preprint articles and online resources [34]. In this study, we have used only posterior-to-anterior (PA) or anterior-to-posterior (AP) CXRs as this view of radiography is widely used by the radiologist.

**QU-COVID-family dataset:** Five major sources were used to create the COVID-family image database: Italian Society of Medical and Interventional Radiology (SIRM) COVID-19 Database [35], Novel Corona Virus 2019 Dataset, Radiopaedia [36], Chest Imaging (Spain) at thread reader and online articles and news-portals [37]. SIRM COVID-19 database [35] shared 94 CXR images from 71 confirmed COVID-19 positive patients in the database. Joseph Paul Cohen *et al.* [38] have created a public database in GitHub by collecting radiographic images of COVID-19, MERS, SARS, and Acute respiratory distress syndrome (ARDS) from the published articles and online resources. 134 COVID-19 positive CXR images were collected from the GitHub database. A physician has shared 103 images from his hospital from Spain to the Chest imaging at thread reader, while 60 images were collected from recently published articles and 32 images were collected from Radiopaedia. The articles, news-portal and online public databases are published from different countries of the World, where COVID-19 has affected significantly and the CXR images, therefore, represent different age groups, gender, and ethnicity from each country.

SARS and MERS CXR images are even scarcer compared to COVID-19; therefore, we collected and indexed CXR images from different publicly available online resources and articles. SARS and MERS radiographic images were collected from 55 different articles (25-MARS, 30-SARS). A total of 260 images was collected from articles and 18 images were collected from Joseph Paul Cohens' GitHub database [38]. Out of 260 images, 70 MERS CXR images were collected from [39], while 16 SARS CXR images were from [40]. During the collection, the authors looked to the peer-reviewed articles in order to ensure the quality of the provided information. Extremely low-resolution images were removed from the database. The collected dataset is highly diverse as CXR images are from several countries around the world and from different X-ray machines. The dataset encapsulates images of different resolution, quality, and SNR levels, as shown in Figure 2.

**Montgomery and Shenzhen CXR lung mask dataset:** This dataset consists of 704 CXR images with their corresponding lung segmentation masks. It was used as initial ground truth masks to train the segmentation model for identifying lung regions. The dataset was acquired by Shenzhen Hospital in China [41], and the Department of Health and Human Services of Montgomery County, MD, USA [42]. Montgomery dataset consists of 80 Normal and 58 unhealthy CXR images with lung segmentation masks, while Shenzhen dataset compromises 326 Normal and 336 unhealthy CXR images. Out of 662 CXRs, 566 CXR images have corresponding ground-truth lung masks.

### *B. **Pre-Processing Techniques***

Medical images are sometimes poor in contrast and often corrupted by noise due to different sources of interference, such as the imaging process and data acquisition. As a result, it may become harder to evaluate them visually. Contrast enhancement methods can play an important role in improving the image quality to provide a better interpretable image to the medical doctors. Besides, it can boost the performance of deep recognition systems. In order to investigate potential enhancement on the classification performance, four different pre-processing schemes were evaluated in this study: original CXR image, which did not undergo any form of pre-processing, contrast limited adaptive histogram equalization (CLAHE), image complementation, and finally, the combination of the three (original, CLAHE, complemented) schemes applied altogether to form a 3-channel approach. Histogram equalization (HE) is a technique mainly used with images that are predominantly dark to enhance the contrast by effectively spreading out the most frequent intensity values [43]. The HE transformation can be defined as follows:

$$y = T(x) = (L-1)\sum_{i=0}^{x} p_x(X = i) \tag{1}$$

where $X$ denotes the random variable representing the original pixel intensities, $p_x(X = x)$ is the probability of having the pixel intensity $x$, $T(x)$ is the transformation function, $y$ are the new intensities after transformation, and $L = 2^N$ is the intensity value for an N-bit image (e.g., for an 8-bit grayscale image, L-1=255 is the maximum intensity value). A closer look at Equation (1) will reveal the fact that $T(x)$ is the approximations of the cumulative distribution function [43]. An improved histogram equalization (HE) variant is called adaptive histogram equalization (AHE). The adaptive equalization performs HE over small regions (patches) in the image. It improves local contrast and edges adaptively in each patch according to the local distribution of pixel intensities instead of the global information of the image. However, AHE could over amplify the noise component in the image [44]. To address this issue, contrast limited adaptive histogram equalization (CLAHE) limits the amount of contrast enhancement that can be produced within the selected region by a threshold parameter. Therefore, produced images are more natural in appearance than those enhanced by AHE [45]. Besides, the clarification of image details is improved. [43].

When the HE technique was applied to the CXR images, it was observed that it saturates some regions. However, CLAHE can overcome this drawback in general. For instance, Figure 3 shows the application of CLAHE and HE techniques over a sample CXR image. The histogram for the equalized images shows that the values are redistributed across all pixels compared with the histogram of the original image. The CLAHE image showed a bell-shaped histogram as Rayleigh distribution was used for the transformation, while the HE showed a flat histogram with a uniform distribution. However, the image was saturated in the center of the lungs when the HE technique was applied. Besides, some regions of the HE image show a sharp brightness difference, whereas the CLAHE image exhibits a smooth transition of intensities for adjacent pixels. As a result, in this study, CLAHE was used for pre-processing the CXR images instead of standard HE.

The image inversion or complement is a technique where the zeros become ones and ones become zeros. Thus, black and white are reversed in a binary image. For an 8-bit greyscale image, the original pixel is subtracted from the highest intensity value, 255, the difference is considered as pixel values for the new image. The mathematical expression is:

$$y = 225 - x \tag{2}$$

where $x$ and y are the intensity values of the original and the transformed (new) images. This technique shows the lungs area (i.e., the region of interest) lighter, and the bones are darker. As this is a standard procedure, which is used widely by radiologists, it may equally help deep networks for a better classification. It can be noted that the histogram for the complemented image is a flipped copy of the original image (Figure 4).

Finally, as shown in Figure 5, the 3-channel scheme was used as the input to the deep networks, where original, CLAHE, and complement images were used altogether. The pixel values for each image are concatenated into a single matrix in order to create a new image. This 3-channel approach is expected to enhance the network performance compared to grayscale CXR images as the utilized deep CNN classifiers were initially pre-trained on RGB images from the ImageNet dataset.

### *C. Chest X-ray Lung Segmentation*

Recently U-Net [46] architecture has gained increasing popularity in different biomedical imaging applications achieving state-of-the-art performance in image segmentation. In this study, original U-Net architecture is utilized to identify the lung regions from the CXR images. U-Net model consists of a contracting path with four encoding blocks, followed by an expanding path with four decoding blocks. Each encoding block consists of two consecutive 3x3 convolutional layers followed by a max-pooling layer with a stride of two for downsampling. In contrast, the decoding blocks consists of a transposed convolutional layer for up-sampling, followed by concatenation with the corresponding feature map from the contracting path, and two subsequent 3x3 convolutional layers. The number of feature channels is doubled at each down-sampling step. In contrast, it is halved for each up-sampling step. All convolutional layers are followed by batch normalization and rectified linear unit (ReLU) activation. At the final layer, 1x1 convolution is utilized to map the output from the last decoding block to two-channel feature maps, where a pixel-wise SoftMax activation function is applied to map each pixel into a binary class of background or lung.

### *D. Chest X-ray Classification*

Transfer learning is a well-established Deep Learning approach, where gained knowledge from one problem is applied to a different but related problem. In this study, four pre-trained CNN models, ResNet18 [47], SqueezeNet [48], InceptionV3 [49], and DenseNet201 [50] were used to classify COVID family CXR images. The deep CNNs were previously trained over the ImageNet database [51]. The rich set of powerful and informative features learned by these networks was utilized through transfer learning to extract specific features of each corona virus-infected lung region. The output layer of each network was replaced by a SoftMax layer with three neurons to classify the X-ray images into one of the following classes: COVID-19, SARS, or MERS.

Overfitting, which is a well-known paradigm for deep networks trained over limited size datasets, can drastically diminish the generalization performance. The problem becomes worse when a high number of training epochs are performed where network saturation would eventually occur due to the vanishing gradient problem, especially at the first hidden layers. With the introduction of the concept of a residual network (ResNet), the vanishing gradient problem with deep CNN networks is solved by introducing the concept of shortcut connections, where the activations of one-layer are fed to the next layer, are fed to the deeper layers as well. ResNet consists of eight residual blocks, where each block has two convolutional layers with 3x3 kernels. The depth of the layers increases every two blocks as going deeper in the network with layer sizes of 64, 128, 256, 512 kernels, respectively. Besides, a 7x7 Convolution layer followed by a pooling layer of stride 2 is used at the start and a SoftMax classification layer at the end of the network.

InceptionV3 showed improved performance in classifying different types of problems. Typically, larger kernels are favored for global features that are distributed over a large area of images, while smaller kernels are preferred for an area-specific feature that is distributed over an image frame. This inspired the idea of inception layers, where kernels of different sizes (1x1, 3x3, and 5x5) are concatenated within the same layer instead of going deeper in the network. The Inception network starts with multiple conventional layers of 3x3 kernel followed by three inception blocks and ends with an 8x8 global average-pooling layer followed by SoftMax classifier. This architecture increases the network space, where the best features can be selected by training.

SqueezeNet is the smallest network considered in this study with 18 layers only and almost 1.24 million parameters compared to 11.7, 20, and 23.9 million parameters for Resent18, InceptionV3, and DenseNet201, respectively. SqueezeNet introduces fire modules, where a squeeze convolutional layer with 1x1 kernels is fed to an expand layer that has a mix of 3x3 and 1x1 kernels. The network begins with a standalone convolutional layer, followed by eight fire blocks, and ends with a convolutional layer followed by a SoftMax layer. The number of kernels per fire module is increased gradually through the network. The network performs max-pooling operation after the first convolutional layer, fourth fire module, and eighth fire module. The compact architecture of SqueezeNet makes it favorable over other networks for such problems that it can achieve comparable performance.

Unlike the residual networks, DenseNet concatenates all feature maps instead of summing residuals. All layers in a dense block are densely connected to their subsequent layers, receiving more supervision from previous layers. This will create compact layers with little redundancy in the learned feature, where dense layers can share pieces of collective knowledge. DenseNet201 consists of four dense blocks, where each block consists of multiple convolution layers with 1x1 and 3x3 filters. The dense blocks are separated by transition layers consisting of batch normalization layer, 1x1 convolutional layer, and 2x2 average pooling layer. The network starts with a 7x7 convolutional layer followed by a 3x3 max-pooling layer, both with a stride of 2, and ends with a 7x7 global average-pooling layer followed by a SoftMax layer.

Choosing the best network for a specific problem is usually a tradeoff between the following two criteria: computational complexity and classification accuracy. Therefore, it is important to investigate different networks to identify the best performing structure for a targeted problem. Since COVID family CXR images have been investigated for the first time in this work, shallow and deep networks with sequential, residual, and dense connections were investigated to identify the best performing one.

### *E. Visualization using Score-CAM*

Visualization techniques help in understanding the internal mechanisms of CNN and the reasoning behind the network making a specific decision. Besides, it interprets the results in a way that is easily understandable to a human, thereby increasing the confidence of CNN outcomes. The likelihood map of pathologies can be generated from deep CNN classifiers by utilizing visualization techniques. In general, locations with peak values in the heat map corresponds to the presence of the disease pattern with a high probability. The main visualization technique employed in literature is Gradient-weighted class Activation Map (Grad-CAM) [52], where activation maps are generated by backward passing the gradients of the target class back to the final convolutional layer in the network to produce the localization map. The localization map Grad-CAM $L^{c}_{Grad-CAM} \in \mathbb{R}^{hxw}$ of height $h$ and width $w$ for class $c$ is obtained by first computing the gradients of the score of the target class with respect to the feature map $A^k$ as $\frac{\partial y^c}{\partial A^k}$ where $y^c$ is the network output before

SoftMax. Next, the gradients are backward passed through global average pooling to compute the $\alpha$ weights, which highlights the importance of feature map k for the decision making of target class c:

$$\alpha_k^c = \frac{1}{Z}\sum_i \sum_j \frac{\partial y^c}{\partial A_{ij}^k} \tag{3}$$

Finally, a weighted combination of activation maps $A^k$ is followed by ReLU to obtain Grad-CAM map:

$$L^C = ReLU(\sum_k \alpha_k^c A^k) \tag{4}$$

Recently, Score-CAM [52] was proposed as a promising alternative to GRAD-CAM. Score-CAM gets rid of the dependencies on gradients by obtaining the weight of each activation map through forward passing scores of the target class. Given a CNN model $y^c = f(CXR)$ that takes an input $CXR$ image and outputs a scalar, $y^c$. The contribution of a specific feature map $A^k$ toward output $y^c$ is defined as follows:

$$\alpha_k^c = f\left(CXR \circ D_l^k\right) - f(CXR) \tag{5}$$

where

$$D_l^k = n(Up(A_l^k)) \tag{6}$$

$Up(.)$ denotes the up-sampling operation of $A$ into the input $(CXR)$ size, $n(.)$ is a normalization function that maps elements of the input matrix into [0, 1], and $\circ$ is the element-wise multiplication. Finally, the Score-CAM saliency map is computed using the same equation as Grad-CAM (Equation (4)). In this study, the Score-CAM method was deployed to visualize the classification outputs of the proposed COVID family recognition system.

***F. Experimental Setup***

The proposed framework is a 2-stage image recognition system using the concatenation of lung segmentation and classification networks. The U-Net segmentation network was pre-trained and validated on the Montgomery and Shenzhen dataset [41, 42], which consists of 704 CXR images with their corresponding lung masks. The pre-trained U-Net model was used to create lung masks for COVID-19, MERS, and SARS CXR images. The masks created by the trained U-Net model were qualitatively assessed to evaluate the performance of the segmentation model. Lung masks were fine-tuned by the MDs to develop ground truth masks for the QU-COVID-family dataset, which will be released as the ground truth masks for the QU-COVID-family dataset. The deep classification networks were evaluated on the compiled QU-COVID-family dataset and their corresponding lung masks. Two classification schemes were considered: plain CXR classification and segmented CXR classification.

Both segmentation and classification networks were trained using 5-fold cross-validation (CV), with 80% train and 20% test (unseen folds), where 20% of training data is used as a validation set to avoid overfitting. Class imbalance in the dataset has a major impact on the model performance of Deep Learning classification problems. Therefore, we balanced the size of each class in the train set using data augmentation. We performed data augmentation by applying rotations of 5, 10, 20, 25, and 30 degrees and horizontal and vertical image translations within the interval [-0.15, +0.15] for the SARS and MERS images. Table 1 summarizes the number of images per class used for training, validation, and testing at each fold. U-Net segmentation model was implemented using PyTorch library with Python 3.7 while MATLAB 2020a was used to train and evaluate the deep CNN classification networks (SqueezeNet, ResNet18, InceptionV3, and DenseNet201); on Intel® Xeon® CPU E5-2697 v4 @2.30GHz and 64 GB RAM, with an 8-GB NVIDIA GeForce GTX 1080 GPU card. U-Net model was trained using Adam optimizer with learning rate, $\alpha = 10^{-3}$, momentum updates, $\beta_1 = 0.9\ and\ \beta_2 = 0.999$, mini-batch size of 4 images with 50 backpropagation epochs. On the other hand, classification CNN models were trained using stochastic gradient descent (SDG) with momentum optimizer, with learning rate, $\alpha = 10^{-3}$, momentum update, $\beta = 0.9$ and mini-batch size of 4 images with 10-20 Back Propagation epochs, as shown in Table 2. Early stopping criterion of six maximum epochs when no improvement in validation loss is seen was used for both the case. Fivefold cross-validation results were averaged to produce the final receiver operating characteristic (ROC) curve, confusion matrix, and evaluation metrics.

The performance of different CNNs was assessed using different evaluation metrics with 95% confidence intervals (CIs). Accordingly, CI for each evaluation metric was computed as follows:

$$r = z\sqrt{metric(1 - metric)/N} \tag{7}$$

where, N is the number of test samples, and $z$ is the level of significance that is 1.96 for 95% CI.

*Evaluation Metrics for Segmentation Model*

The performance of the U-Net segmentation network was evaluated on pixel-level, where the foreground (lung region) was considered as the positive class, and background as the negative class. Three evaluation metrics were computed to evaluate the segmentation performance:

$$Accuracy = \frac{TP + TN}{TP + TN + FP + FN} \tag{8}$$

where $accuracy$ is the ratio of the correctly classified pixels among the image pixels. TP, TN, FP, FN represent the true positive, true negative, false positive, and false negative respectively.

$$Intersection\ over\ Union\ (IoU) = \frac{TP}{TP + FP + FN} \tag{9}$$

$$Dice\ Similarity\ Coefficient\ (DSC) = \frac{2 * TP}{2 * TP + FP + FN} \tag{10}$$

where, both $IoU\ and\ DSC$ is a statistical measure of spatial overlap between the binary ground-truth segmentation mask and the predicted segmentation mask, whereas the main difference is that DSC considers double weight for $TP$ pixels (true lung predictions) compared to IoU.

*Evaluation Metrics for Classification Models*

Five evaluation metrics were considered for the classification scheme: accuracy, sensitivity, precision, f1-score, and specificity. Per-class values were computed over the overall confusion matrix that accumulates all test fold results of the 5-fold cross-validation.

$$Accuracy_{class_i} = \frac{TP_{class_i} + TN_{class_i}}{TP_{class_i} + TN_{class_i} + FP_{class_i} + FN_{class_i}} \tag{11}$$

where $accuracy$ is the ratio of correctly classified CXR samples among all the data.

$$Precision_{class_i} = \frac{TP_{class_i}}{TP_{class_i} + FP_{class_i}} \tag{12}$$

where $precision$ is the rate of correctly classified positive class CXR samples among all the samples classified as positive samples.

$$Sensitivity_{class_i} = \frac{TP_{class_i}}{TP_{class_i} + FN_{class_i}} \tag{13}$$

where $sensitivity$ is the rate of correctly predicted positive samples in the positive class samples,

$$F1_score_{class_i} = 2\frac{Precision_{class_i} \times Sensitivity_{class_i}}{Precision_{class_i} + Sensitivity_{class_i}} \tag{14}$$

where $f1_score$ is the naïve average of precision and sensitivity.

$$Specificity_{class_i} = \frac{TN_{class_i}}{TN_{class_i} + FP_{class_i}} \tag{15}$$

where specificity is the ratio of accurately predicted negative class samples to all negative class samples. Besides $class_i = COVID_19, MERS\ or\ SARS$.

The overall performance was computed using the weighted average values of each class. The weighted average gives a better estimation of the overall performance as class frequencies vary for the presented problem.

$$metric_x = \frac{n1(metric_{x_class_1}) + n2(metric_{x_class_2}) + n3(metric_{x_class_3})}{n1 + n2 + n3} \quad (16)$$

where $metric_x = \text{Accuracy}, \text{Precision}, \text{Sensitivity}, \text{F1}_{\text{score}}, \text{or specificity}.$ Finally, $class_1, class_2,\ \ and\ class_3$ are COVID-19, MERS, and SARS with $n1, n2, and\ n3$ samples, respectively.

In the segmentation task, we aimed to minimize the false negatives (FNs), which are the misclassified lung pixels, and to minimize the false positives (FPs), the background pixels, which are misclassified as lung areas to ensure that the network learns from the exact region of interest for the subsequent classification task. Equivalently, we aim to maximize the DSC and IoU values. The main objective of our recognition scheme is to maximize per class sensitivities for COVID-19, MERS, and SARs classes, especially for the COVID-19 class. This ensures reliable medical diagnosis, which can save time, resources and help in delivering the right treatments for patients in the current pandemic.

## III. Results & Discussion

The performance of the proposed 2-stage image recognition system is detailed in this section. The deep CNN-based classification networks were evaluated on the CXR images of the compiled QU-COVID-family dataset. Two classification schemes (plain and segmented CXR classification) were evaluated, and the outcome was interpreted with the help of Score-CAM visualization technique.

The U-Net segmentation model was trained and evaluated on 704 CXR samples with ground-truth lung masks of the Montgomery and Shenzhen dataset as shown in Table 3. The model showed promising segmentation performance with IoU and DSC of 93.11% and 96.35%, respectively, on the two publicly available datasets. The qualitative evaluation of the trained U-Net model on the compiled QU-COVID-Family dataset is presented in Figure 6. The model can reliably segment the lung images if the lung areas are distinguishable. However, the segmentation network suffers from severely infected lungs due to the whitened infection area in the lungs. Predicted lung masks by the U-Net model were revised by medical doctors to ensure that the segmentation masks encapsulate the entire lung region. This ensures that neither important lung areas, such as peripheral parts with infection, are discarded, nor non-lung regions, such as heart and spines, are included within the lung mask.

Table 4 summarizes the classification performances of the deep CNN models in-terms of the per-class performance metrics for plain and segmented X-ray image classifications. For each network, four different pre-processing schemes (original, CLAHE, complemented, and 3-channel) were compared, and the best performing scheme is presented. For plain X-ray images, it was observed that SqueezeNet achieved the best classification performance on original images, while ResNet18 and Inceptionv3 outperformed on 3-channel images. For the segmented lung X-ray images, SqueezeNet and InceptionV3 showed the best performance with the original lung images without any pre-processing, and InceptionV3 outperformed all the networks. On the other hand, ResNet18 and DenseNet201 performed better on 3-Channel images. In general, the investigated CNN models showed high COVID-19 sensitivity values (>96%) for segmented data, while it showed varying results with plain X-rays. For instance, with plain X-ray, SqueezeNet showed 91.97% COVID-19 sensitivity, while InceptionV3 showed 99.53% COVID-19 sensitivity. For SARS and MERS cases, the InceptionV3 network achieved the highest sensitivities for plain and segmented lung X-ray images. The sensitivity for MERS and SARS detection were 93.1%/79.68% and 97.04%/90.26% for plain/segmented lung CXRs, respectively. It is evident that the overall performance for MERS detection significantly degrades for the segmented lung images. This is most likely due to a large number of lower quality CXR images in the MERS dataset. Even though the performance degrades with segmented lungs, as the network learns from the main region of interest (lung area), the results obtained from the segmented lungs are much more reliable.

Figure 7 shows the comparative ROC curves for different networks for different pre-processing schemes with plain and segmented X-rays. For plain X-rays, it is apparent from Figure 7(A) that Inceptionv3 outperforms other models over the original dataset while DenseNet201 and ResNet18 obtain a close performance, even though DenseNet201 is a very deep network compared to ResNet18. In contrast, the performance of SqueezeNet is comparable to the significantly deeper network, DenseNet201. Interestingly, the performances of InceptionV3, ResNet18, and DenseNet201 are comparable in the case of CLAHE images, and SqueezeNet shows a promising performance as well (Figure 7 (B)). However, there is no notable performance improvement observed by this pre-processing scheme rather than making the classification less network independent. Figure 7(C) shows that significant performance improvement can be achieved using deeper networks with the complemented image. In contrast, the performance degrades for ResNet18 and especially for SqueezeNet. Figure 7(D) clearly depicts that the 3-channel scheme significantly improves the classification performance of InceptionV3 and ResNet18. However, this is not the case for DenseNet201 and SqueezeNet. InceptionV3 using a 3-channel scheme achieved the overall best classification performance among the four networks. Interestingly, these four pre-trained networks showed similar comparative accuracies while evaluated on the ImageNet database [53]. On the other hand, with segmented X-rays, the four networks showed close performance for different pre-processing techniques (Figure 7(E-H)). Therefore, it can be concluded that proper segmentation can guide the network to learn from lung regions mainly. Thus, it makes the classification problem less dependent on the pre-processing technique. In addition, it eases the recognition task for the shallow networks, allowing them to achieve comparable results to their deeper competitors. Consequently, InceptionV3 using the original dataset without any pre-processing, showed the best classification performance for segmented X-rays.

Since InceptionV3 is the top-performing network, we used this network to investigate the role of different pre-processing schemes on the classification performance (Table 5). An interesting observation for the plain X-rays is that the network's performance has been significantly dropped for the images enhanced with the CLAHE technique, while the complement and 3-channel approach showed a significant boost in the network performance. However, image enhancement techniques failed to provide any boost in the network performance for the segmented X-rays images. This is most likely because the image complement technique was providing a performance boost for the plain X-ray images. Thus, the 3-channel approach was boosted. On contrary, for the segmented lungs, image complement was not showing superior performance compared to the original image. Therefore, the 3-channel approach did not add any gain in the process. In a nutshell, the performance gain from a specific pre-processing technique is both problem and network dependent. Additionally, for our future work, it is worth investigating the effect of the ensemble technique on the X-ray classification scheme. The ensemble approach combines the output from several networks trained with different pre-processing techniques to generate the final classification output. This is different from the 3-channel scheme used in this study, where the variants of the pre-processed X-ray are combined and fed to a single network to make the final decision.

Figure 8 shows the overall confusion matrix cumulated from all folds of InceptionV3 network for plain and segmented lung X-ray images. It is apparent from Figure 8 that the performance of the InceptionV3 has degraded by the use of segmentation, and more images were miss-classified. However, the network is restricted to learn from the lung areas, as confirmed by the Score-CAM saliency map (Figure 9). In contrast, with plain X-rays, the network is learning irrelevant features from non-lung regions, such as the heart, bones, or background. Figure 10 shows sample miss-classified X-ray images, corresponding lung image and Score-CAM visualization for COVID-19, MERS, and SARS images to identify the potential reasons of the network failure. It can be seen from Figure 10 that InceptionV3 failed to classify the lung images properly if the network did not learn from the lung areas exclusively. On the other hand, for those images that are correctly classified by the network, Score-CAM is showing that the CNN model is learning from the entire lung region. Therefore, it can be summarized that the reliable segmentation of lungs from the X-ray images and the use of segmented X-ray images for the classification problem can significantly boost the reliability and performance of AI-based solutions for computer-aided-diagnostic applications.

## IV. Conclusion

In this study, we aim to investigate if it is possible to discriminate CoV family (COVID-19, MERS, and SARS) infestations directly from CXR images. For this purpose, we trained and evaluated several state-of-the-art Deep Learning networks. To accomplish this objective, we compiled a unique dataset, so-called QU-COIVD-family that will be released along with this study as a benchmark dataset in this domain. The dataset encapsulating 701 X-ray images from numerous countries (e.g. Italy, Spain, China, etc.) and different X-ray machines with varying quality, resolution, and noise levels. Due to the scarcity of the data, we have used the transfer-learning paradigm with certain data augmentation and several pre-processing schemes to improve the classification performance and robustness. We proposed a cascaded COVID family recognition system, where first lung regions are identified using the U-Net model, and then a deep CNN classifier (SqueezeNet, ResNet18, DenseNet201 or InceptionV3) is used to classify the patient as COIVD-19, MERS, or SARS. The proposed system was evaluated on two recognition schemes: plain X-ray classification and segmented X-ray classification. InceptionV3 model with 3-channel data yields the best performance for the first scheme with 98.2% accuracy, 97.8% precision, 97.7% sensitivity, 97.8% f1-score, and 97.1% specificity. In contrast, InceptionV3 with segmented lungs from the original X-ray images showed the best results with 92.2% accuracy, 92.1% precision, 92.1% sensitivity, 92.1% f1-score, and 93.8% specificity. Even though segmented CXR showed lower performance compared to plain X-ray classification, their results are more reliable and trustworthy as the network is restricted to learn only from the lung areas. Furthermore, all networks showed high COVID-19 sensitivity (>96%) with segmented lung images. This indicates the unique radiographic signature of COVID-19 cases in the eyes of AI, which is often a challenging task for medical doctors. Besides, comparing the achieved results with recent literature as shown in Table 6, our proposed model performed very well. Moreover, several interesting observations can be made from the obtained results. Firstly, the proposed pre-processing schemes can be useful particularly for some networks and can improve network performance significantly. In particular, the 3-channel scheme yielded the overall best performance level for the deep CNN, InceptionV3; however, it cannot be generalized for all networks. This shows that the performance gain from a particular pre-processing scheme is both network and problem dependent. A close look at the successful and failed cases reveals the fact that CNN was learning predominantly from the segmented lung X-ray images for the successful cases while that was not the case for the miss-classified images. Therefore, lung segmentation along with Score-CAM based visualization technique can significantly help to produce reliable performance from the Deep Learning models in the computer-aided-diagnostic applications.

## Acknowledgment

This work was made possible by Qatar University COVID19 Emergency Response Grant (QUERG-CENG-2020-1). The statements made herein are solely the responsibility of the authors.

## Data availability

Chest X-ray images database for COVID-19, MERS and SARS is available online at www.kaggle.com/dataset/057e1b6dc41d9691e59dded4445fa8cc2f0b4b5cbcb49aef9583d95233799d5a (Private link, the database will be available for the public once the paper is accepted)

## Compliance with Ethical Standards

**Conflict of Interest:** The authors declare that they have no conflict of interest.

**Ethical Approval:** This article uses the X-ray images, which was made publicly available by different research groups as mentioned in Section II. Therefore, the authors of this study were not involved directly with human participants or animals.

# Tables

Table 1. Number of images per class and per fold before and after data augmentation for CXR lung segmentation and CXR classification tasks.

| TASK | DATASET | CLASS | # OF SAMPLES | TRAINING SAMPLES | AUGMENTED TRAINING SAMPLES | VALIDATION SAMPLES | TEST SAMPLES |
|---|---|---|---|---|---|---|---|
| CXR Lung Segmentation | Montgomery and Shenzhen dataset [40, 41] | CXRs with corresponding lung masks | 704 | 452 | 1808 | 112 | 140 |
| CXR Classification | QU-COIVD-family | COVID-19 | 423 | 270 | 1890 | 68 | 85 |
| | | MERS | 144 | 92 | 1932 | 23 | 29 |
| | | SARS | 134 | 89 | 1806 | 21 | 26 |

Table 2. Details of segmentation and classification models training parameters

| TRAINING PARAMETERS | SEGMENTATION MODEL | CLASSIFICATION MODEL |
|---|---|---|
| batch size | 4 | 4 |
| learning rate | 0.001 | 0.001 |
| learning rate drop factor | 0.1 | 0.1 |
| max epochs | 50 | 20 |
| epochs patience | 3 | 3 |
| epochs stopping criteria | 6 | 6 |
| optimizer | Adam | Stochastic Gradient Descent (SGD) |

Table 3. CXR lung segmentation results (%)

| NETWORK | ACCURACY | IOU | DSC |
|---|---|---|---|
| U-NET | 98.21 ± 0.98 | 93.11 ± 1.87 | 96.35 ± 1.39 |

Table 4. Comparison between four classification networks: SqueezeNet, ResNet18, InceptionV3 and DenseNet201, in terms of Accuracy, Precision, Sensitivity, F1-score and Specificity (%). The best preprocessing technique is reported for each network.

| | NETWORK | CLASS | ACCURACY | PRECISION | SENSITIVITY | F1-SCORE | SPECIFICITY |
|---|---|---|---|---|---|---|---|
| Plain X-rays | SqueezeNet (Original) | COVID-19 | 88.27 ± 3.07 | 89.31 ± 2.94 | 91.97 ± 2.59 | 90.48 ± 2.8 | 82.63 ± 3.61 |
| | | MERS | 91.56 ± 4.54 | 84.97 ± 5.84 | 72.09 ± 7.33 | 77.58 ± 6.81 | 96.58 ± 2.97 |
| | | SARS | 91.86 ± 4.63 | 77.32 ± 7.09 | 81.25 ± 6.61 | 78.9 ± 6.91 | 94.36 ± 3.91 |
| | | Overall | 89.77 ± 2.24 | 86.13 ± 2.56 | 85.84 ± 2.58 | 85.98 ± 2.57 | 88.02 ± 2.4 |
| | ResNet18 (3-Channel) | COVID-19 | 94.04 ± 2.26 | 92.99 ± 2.43 | 97.88 ± 1.37 | 95.29 ± 2.02 | 88.21 ± 3.07 |
| | | MERS | 96.03 ± 3.19 | 94.34 ± 3.77 | 85.49 ± 5.75 | 89.5 ± 5.01 | 98.75 ± 1.81 |
| | | SARS | 97.16 ± 2.81 | 96.17 ± 3.25 | 88.89 ± 5.32 | 91.97 ± 4.6 | 99.12 ± 1.58 |
| | | Overall | 95.02 ± 1.61 | 93.88 ± 1.77 | 93.61 ± 1.81 | 93.74 ± 1.79 | 92.41 ± 1.96 |
| | Inceptionv3 (3-Channel) | COVID-19 | 97.87 ± 1.38 | 97.13 ± 1.59 | **99.53 ± 0.65** | 98.29 ± 1.24 | 95.36 ± 2 |
| | | MERS | 98.3 ± 2.11 | 98.4 ± 2.05 | 93.1 ± 4.14 | 95.56 ± 3.36 | 99.64 ± 0.98 |
| | | SARS | 99.29 ± 1.42 | 99.2 ± 1.51 | 97.04 ± 2.87 | 98.08 ± 2.32 | 99.82 ± 0.72 |
| | | Overall | **98.22 ± 0.98** | 97.79 ± 1.09 | **97.73 ± 1.1** | 97.76 ± 1.1 | 97.07 ± 1.25 |
| | DenseNet201 (complement) | COVID-19 | 96.17 ± 1.83 | 96.55 ± 1.74 | 97.18 ± 1.58 | 96.85 ± 1.66 | 94.64 ± 2.15 |
| | | MERS | 97.02 ± 2.78 | 93.57 ± 4.01 | 91.72 ± 4.5 | 92.63 ± 4.27 | 98.39 ± 2.06 |
| | | SARS | 98.86 ± 1.8 | 97.23 ± 2.78 | 97.04 ± 2.87 | 97.05 ± 2.86 | 99.3 ± 1.41 |
| | | Overall | 96.84 ± 1.29 | 96.07 ± 1.44 | 96.03 ± 1.45 | 96.05 ± 1.44 | 96.28 ± 1.4 |
| Segmented X-rays | SqueezeNet (Original) | COVID-19 | 92.12 ± 2.57 | 91.51 ± 2.66 | 96.22 ± 1.82 | 93.71 ± 2.31 | 85.83 ± 3.32 |
| | | MERS | 91.26 ± 4.61 | 83.01 ± 6.13 | 71.31 ± 7.39 | 75.92 ± 6.98 | 96.41 ± 3.04 |
| | | SARS | 92.88 ± 4.35 | 82.58 ± 6.42 | 80.6 ± 6.7 | 81.28 ± 6.6 | 95.78 ± 3.4 |
| | | Overall | 88.13 ± 2.39 | 88.05 ± 2.4 | 88.13 ± 2.39 | 88.09 ± 2.4 | 89.89 ± 2.23 |
| | ResNet18 (3-Channel) | COVID-19 | 93.01 ± 2.43 | 91.74 ± 2.62 | 97.16 ± 1.58 | 94.37 ± 2.2 | 86.69 ± 3.24 |
| | | MERS | 92.44 ± 4.32 | 85.27 ± 5.79 | 76.39 ± 6.94 | 80.59 ± 6.46 | 96.59 ± 2.96 |
| | | SARS | 95.44 ± 3.53 | 91.13 ± 4.81 | 84.33 ± 6.16 | 87.6 ± 5.58 | 98.06 ± 2.34 |
| | | Overall | 91.12 ± 2.11 | 91.2 ± 2.1 | 91 ± 2.12 | 91 ± 2.12 | 93.58 ± 1.81 |
| | Inceptionv3 (Original) | COVID-19 | 94.84 ± 2.11 | 94.85 ± 2.11 | 96.94 ± 1.64 | 95.82 ± 1.91 | 91.63 ± 2.64 |
| | | MERS | 93.41 ± 4.05 | 86.87 ± 5.52 | 79.68 ± 6.57 | 82.62 ± 6.19 | 96.95 ± 2.81 |
| | | SARS | 96 ± 3.32 | 88.97 ± 5.3 | 90.26 ± 5.02 | 89.58 ± 5.17 | 97.35 ± 2.72 |
| | | Overall | **92.12 ± 1.99** | 92.08 ± 2 | **92.12 ± 1.99** | 92.1 ± 2 | 93.81 ± 1.78 |
| | DenseNet201 (3-Channel) | COVID-19 | 94.12 ± 2.24 | 93.2 ± 2.4 | **97.64 ± 1.45** | 95.3 ± 2.02 | 88.74 ± 3.01 |
| | | MERS | 93.27 ± 4.09 | 89.75 ± 4.95 | 75.57 ± 7.02 | 81.93 ± 6.28 | 97.84 ± 2.37 |
| | | SARS | 94.86 ± 3.74 | 86.38 ± 5.81 | 87.21 ± 5.65 | 86.42 ± 5.8 | 96.66 ± 3.04 |
| | | Overall | 91.12 ± 2.11 | 91.18 ± 2.1 | 91.12 ± 2.11 | 91.15 ± 2.1 | 92.11 ± 2 |

Table 5. Comparison between different pre-processing schemes for the best performing network for COVID-19 recognition, InceptionV3, in terms of Accuracy, Precision, Sensitivity, F1-score and Specificity (%)

| | Preprocessing | Class | Accuracy | Precision | Sensitivity | F1-score | Specificity |
|---|---|---|---|---|---|---|---|
| Plain X-rays | Original | COVID-19 | 92.86 ± 2.45 | 93.68 ± 2.32 | 94.54 ± 2.17 | 94.08 ± 2.25 | 90.28 ± 2.82 |
| | | MERS | 94.86 ± 3.61 | 88.48 ± 5.21 | 86.08 ± 5.65 | 87.24 ± 5.45 | 97.12 ± 2.73 |
| | | SARS | 96.57 ± 3.08 | 91.25 ± 4.78 | 91.03 ± 4.84 | 90.98 ± 4.85 | 97.88 ± 2.44 |
| | | Overall | 93.87 ± 1.78 | 92.15 ± 1.99 | 92.13 ± 1.99 | 92.14 ± 1.99 | 92.93 ± 1.9 |
| | CLAHE | COVID-19 | 90.56 ± 2.79 | 90.17 ± 2.84 | 94.79 ± 2.12 | 92.4 ± 2.53 | 84.11 ± 3.48 |
| | | MERS | 92.85 ± 4.21 | 89.22 ± 5.07 | 74.19 ± 7.15 | 80.72 ± 6.44 | 97.67 ± 2.46 |
| | | SARS | 95.71 ± 3.43 | 88.23 ± 5.46 | 89.54 ± 5.18 | 88.71 ± 5.36 | 97.17 ± 2.81 |
| | | Overall | 92.04 ± 2 | 89.6 ± 2.26 | 89.56 ± 2.26 | 89.58 ± 2.26 | 89.43 ± 2.28 |
| | Image complement | COVID-19 | 96.6 ± 1.73 | 96.49 ± 1.75 | 98.12 ± 1.29 | 97.24 ± 1.56 | 94.28 ± 2.21 |
| | | MERS | 97.59 ± 2.5 | 97.75 ± 2.42 | 90.34 ± 4.83 | 93.66 ± 3.98 | 99.46 ± 1.2 |
| | | SARS | 98.44 ± 2.1 | 95.35 ± 3.57 | 97.04 ± 2.87 | 96.02 ± 3.31 | 98.77 ± 1.87 |
| | | Overall | 97.14 ± 1.23 | 96.53 ± 1.35 | 96.31 ± 1.4 | 96.42 ± 1.38 | 96.18 ± 1.42 |
| | **3-Channel** | COVID-19 | 97.87 ± 1.38 | 97.13 ± 1.59 | **99.53 ± 0.65** | 98.29 ± 1.24 | 95.36 ± 2 |
| | | MERS | 98.3 ± 2.11 | 98.4 ± 2.05 | 93.1 ± 4.14 | 95.56 ± 3.36 | 99.64 ± 0.98 |
| | | SARS | 99.29 ± 1.42 | 99.2 ± 1.51 | 97.04 ± 2.87 | 98.08 ± 2.32 | 99.82 ± 0.72 |
| | | Overall | **98.22 ± 0.98** | **97.79 ± 1.09** | **97.73 ± 1.1** | **97.76 ± 1.1** | **97.07 ± 1.25** |
| Segmented X-rays | **Original** | COVID-19 | 94.84 ± 2.11 | 94.85 ± 2.11 | **96.94 ± 1.64** | 95.82 ± 1.91 | 91.63 ± 2.64 |
| | | MERS | 93.41 ± 4.05 | 86.87 ± 5.52 | 79.68 ± 6.57 | 82.62 ± 6.19 | 96.95 ± 2.81 |
| | | SARS | 96 ± 3.32 | 88.97 ± 5.3 | 90.26 ± 5.02 | 89.58 ± 5.17 | 97.35 ± 2.72 |
| | | Overall | **92.12 ± 1.99** | **92.08 ± 2** | **92.12 ± 1.99** | **92.1 ± 2** | **93.81 ± 1.78** |
| | CLAHE | COVID-19 | 94 ± 2.26 | 94.12 ± 2.24 | 96.23 ± 1.82 | 95.11 ± 2.06 | 90.59 ± 2.78 |
| | | MERS | 93.29 ± 4.09 | 86.37 ± 5.6 | 80.49 ± 6.47 | 83 ± 6.14 | 96.59 ± 2.96 |
| | | SARS | 94.74 ± 3.78 | 87.73 ± 5.56 | 85.81 ± 5.91 | 86.21 ± 5.84 | 96.84 ± 2.96 |
| | | Overall | 91.01 ± 2.12 | 91.3 ± 2.09 | 91.01 ± 2.12 | 91.16 ± 2.1 | 93.02 ± 1.89 |
| | Image complement | COVID-19 | 92.55 ± 2.5 | 92.08 ± 2.57 | 96.22 ± 1.82 | 94.02 ± 2.26 | 86.94 ± 3.21 |
| | | MERS | 93.11 ± 4.14 | 88.7 ± 5.17 | 75.47 ± 7.03 | 80.72 ± 6.44 | 97.66 ± 2.47 |
| | | SARS | 94.86 ± 3.74 | 86.13 ± 5.85 | 87.29 ± 5.64 | 86.58 ± 5.77 | 96.65 ± 3.05 |
| | | Overall | 90.26 ± 2.19 | 90.25 ± 2.2 | 90.26 ± 2.19 | 90.25 ± 2.2 | 90.99 ± 2.12 |
| | 3-Channel | COVID-19 | 94.56 ± 2.16 | 95.03 ± 2.07 | 96.23 ± 1.82 | 95.56 ± 1.96 | 92 ± 2.59 |
| | | MERS | 92.83 ± 4.21 | 84.24 ± 5.95 | 79.68 ± 6.57 | 81.56 ± 6.33 | 96.22 ± 3.11 |
| | | SARS | 94.29 ± 3.93 | 84.78 ± 6.08 | 85.75 ± 5.92 | 85.03 ± 6.04 | 96.3 ± 3.2 |
| | | Overall | 90.84 ± 2.14 | 90.85 ± 2.13 | 90.84 ± 2.14 | 90.85 ± 2.13 | 93.68 ± 1.8 |

Table 6. Comparing the proposed work with recent literature about automatic COVID-19 diagnosis using CXR images, in terms of achieved Accuracy, Specificty and Sensitivty (%)

| Ref. | Dataset | Methodology | Lung Segmentation | Best Network | Accuracy | Specificity | Sensitivity |
|---|---|---|---|---|---|---|---|
| [16] | 25 COVID-19,<br>25 Viral/Bacterial Pneumonia | Classification:<br>COVID-19/<br>Pneumonia | No | Pretrained CNN as a feature extractor:<br>ResNet50 | 95.4 | 93.4 | 97.2 |
| [18] | 224 COVID-19<br>714 Viral/Bacterial Pneumonia<br>504 Normal | Classification:<br>COVID-19/<br>Pneumonia/<br>Normal | No | Pretrained CNN:<br>MobileNetv2 | 96.7 | 96.5 | 98.7 |
| [19] | 358 COVID-19<br>8,066 Viral/Bacterial Pneumonia<br>5,538 Normal | Classification:<br>COVID-19/<br>Pneumonia/<br>Normal | No | Proposed CNN:<br>COVID-Net | 93.3 | N/A | 93.3 |
| [20] | 403 COVID-19<br>721 Normal | Classification:<br>COVID-19/<br>Normal | No | Pretrained CNN:<br>VGG16 | 95 | 97 | 90 |
| [21] | 423 COVID-19<br>1485 Viral Pneumonia<br>1579 Normal | Classification:<br>COVID-19/<br>Viral /<br>Normal | No | Pretrained CNN:<br>DenseNet201 | 97.9 | 97.9 | 97.9 |
| [26] | 180 COVID-19<br>20 Viral-Pneumonia<br>54 Bacterial-Pneumonia<br>57 Tuberculosis<br>191 Normal | Viral pneumonia (including COVID-19)/<br>Bacterial Pneumonia/<br>Tuberculosis/Normal | Yes | Patch-based classification:<br>ResNet-50 | 88.9 | 96.6 | 85.9 |
| **This work** | 423 COVID- 19/144 MERS/<br>134 SARS | Classification:<br>COVID-19/<br>SARS /<br>MERS | Yes | Pretrained CNN:<br>3-Channel Inception V3 | **98.2/ 92.12** | **97.07/ 93.8** | **97.7/ 92.12** |

# Figures

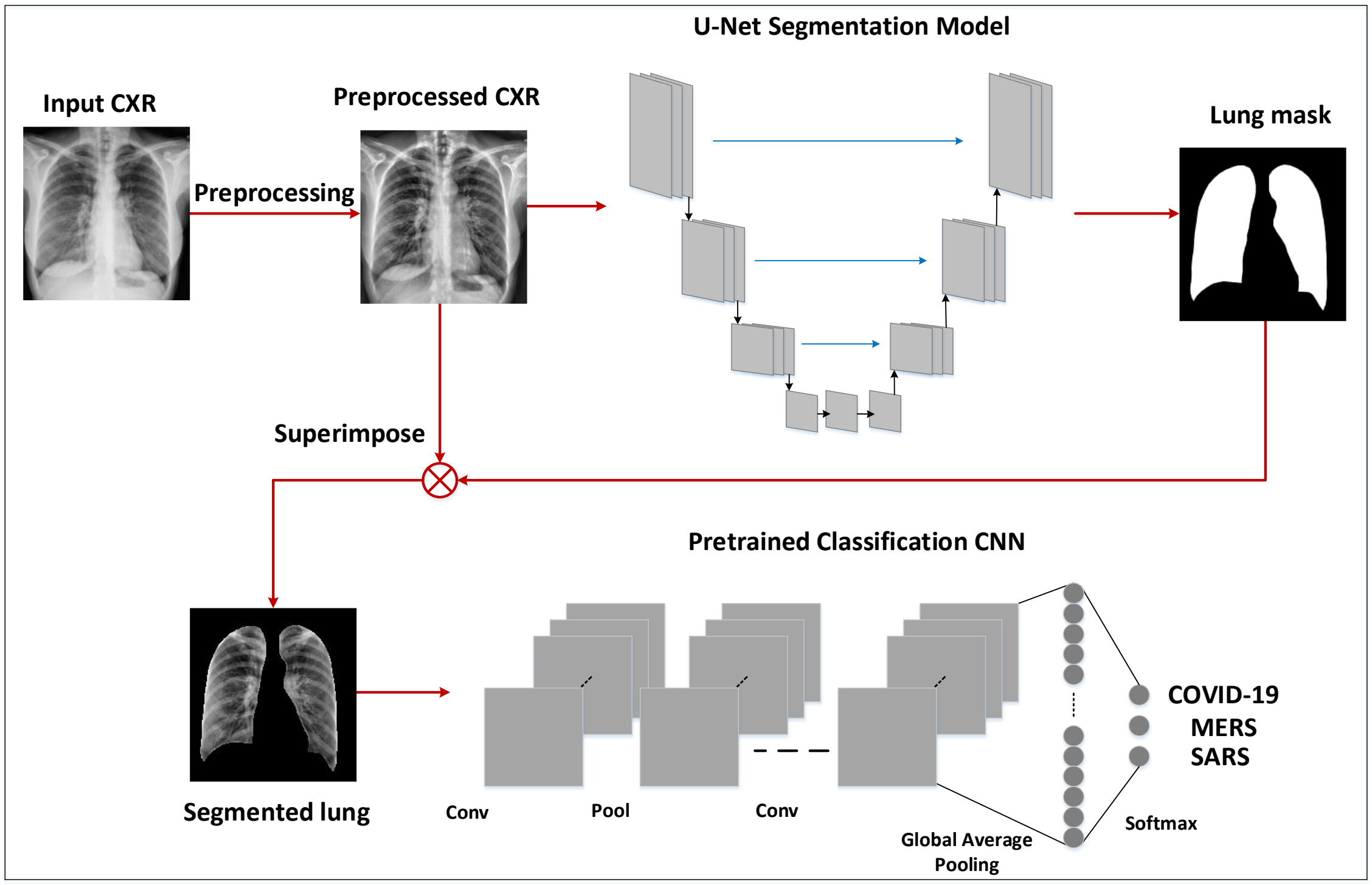

Figure 1. The pipeline of the proposed COVID-19 recognition system. First, the CXR image is pre-processed and segmented using a U-Net segmentation model. Secondly, the segmented CXR is classified using a pre-trained CNN classifier as COVID-19, MERS, or SARS.

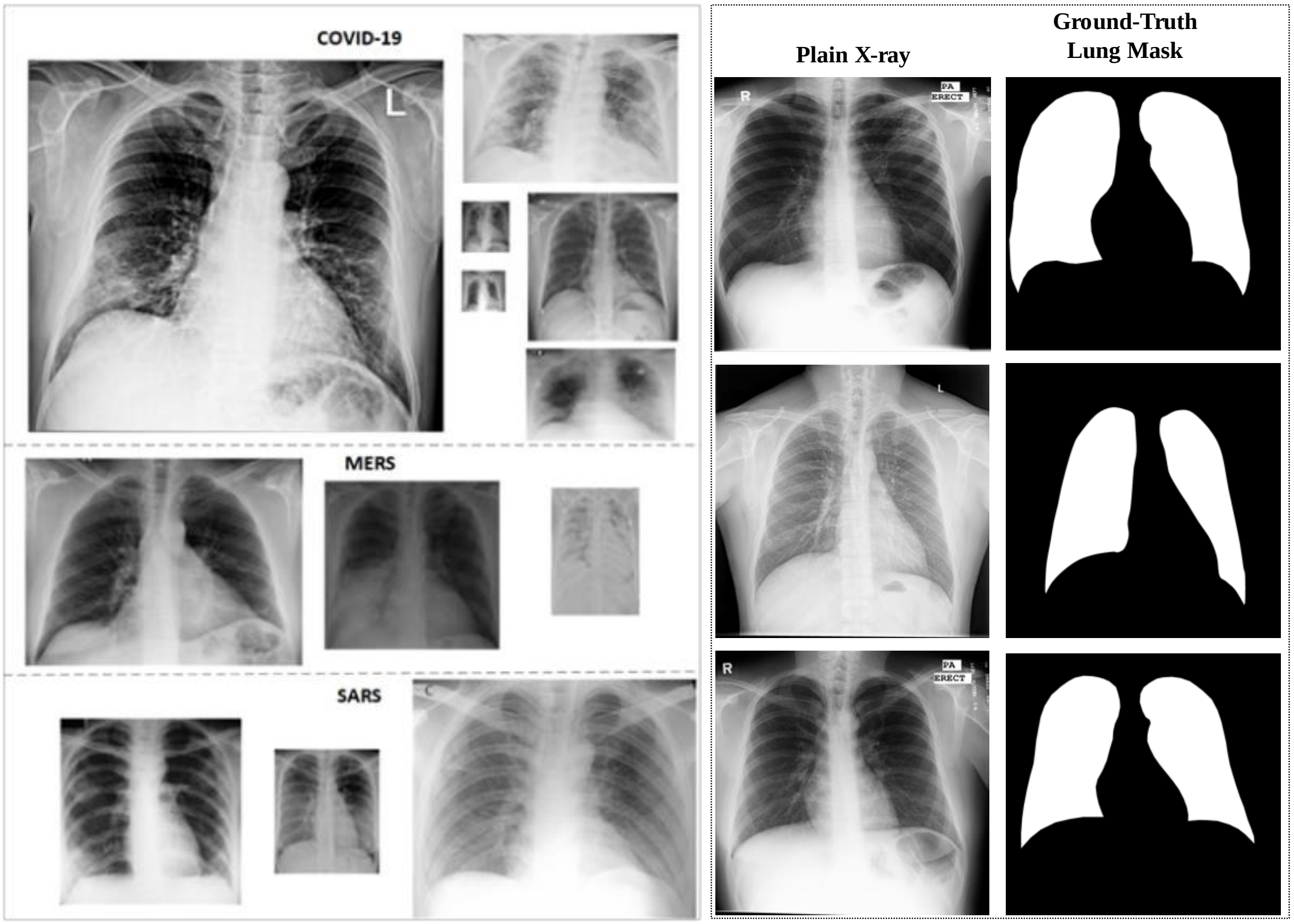

Figure 2. Sample X-ray images from the dataset (left): (A) COVID-19, (B) MERS, and (C) SARS. The dataset encapsulates images from different countries around the world with different resolution, quality, and SNR levels. All images are rescaled with the same factor to illustrate the diversity of the dataset. Sample X-ray images and corresponding ground-truth masks from the segmentation database (right).

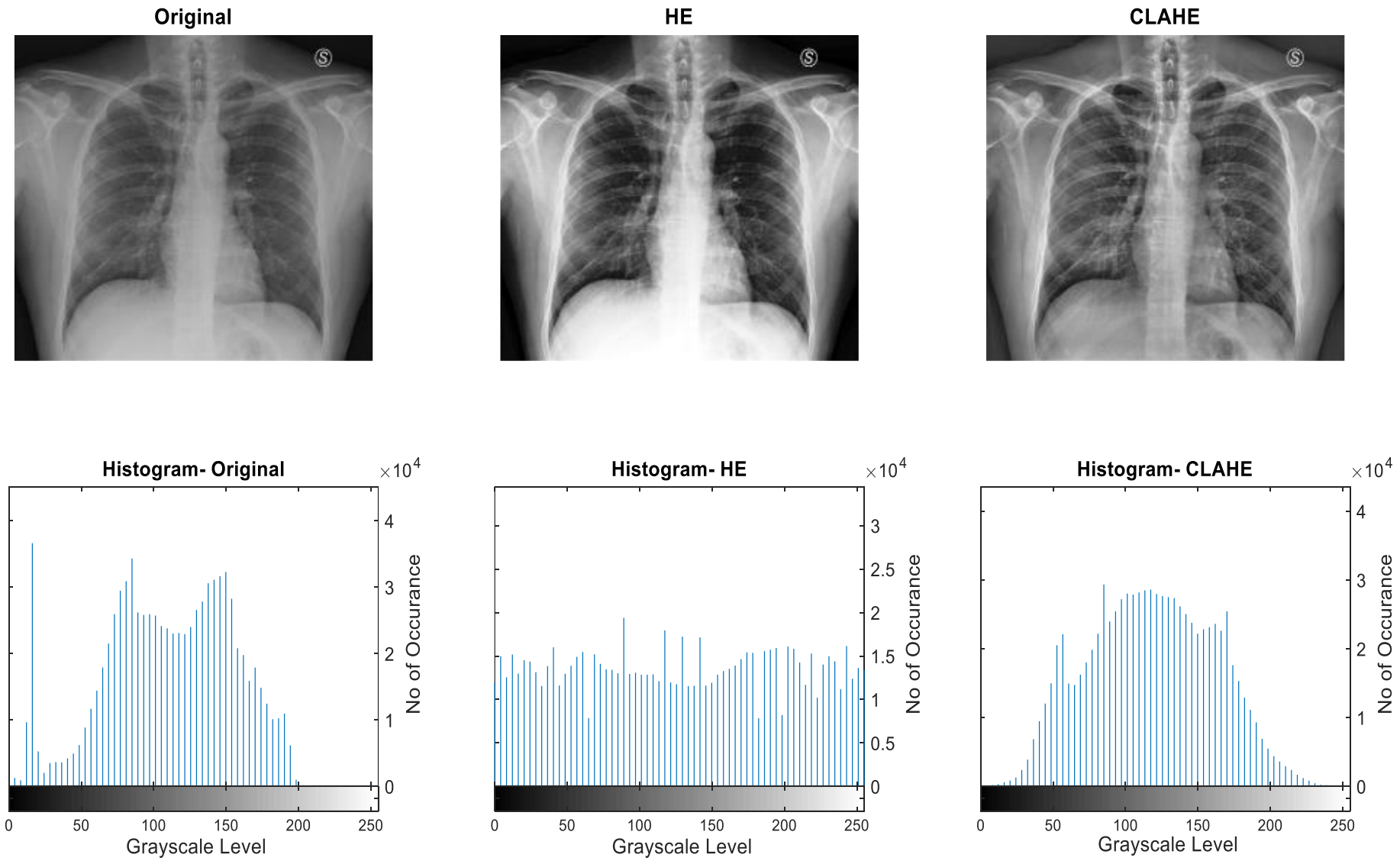

Figure 3. Comparison between original, HE and CLAHE equalized X-ray images with corresponding histograms.

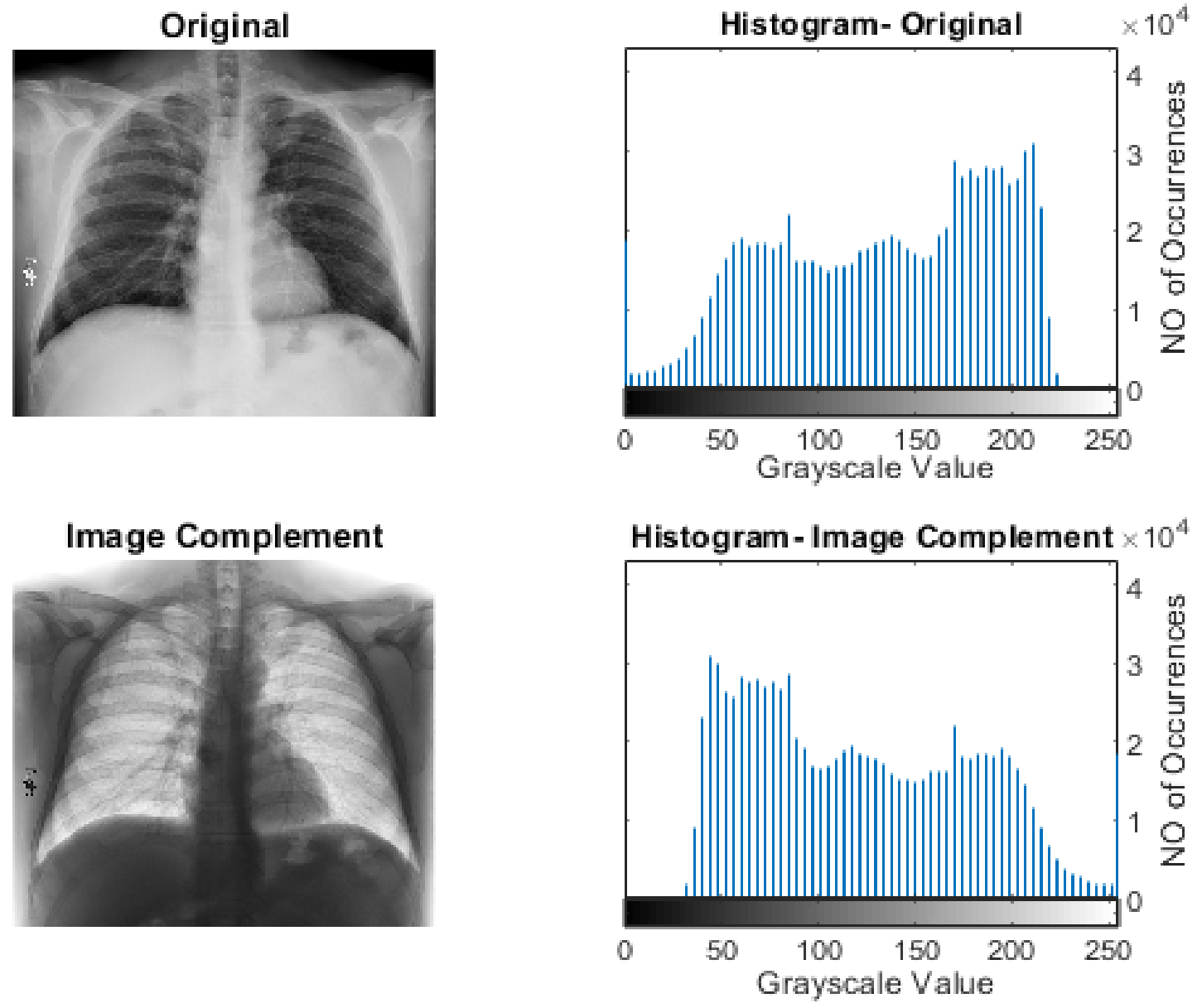

Figure 4. Comparison between an original X-ray and its image complement.

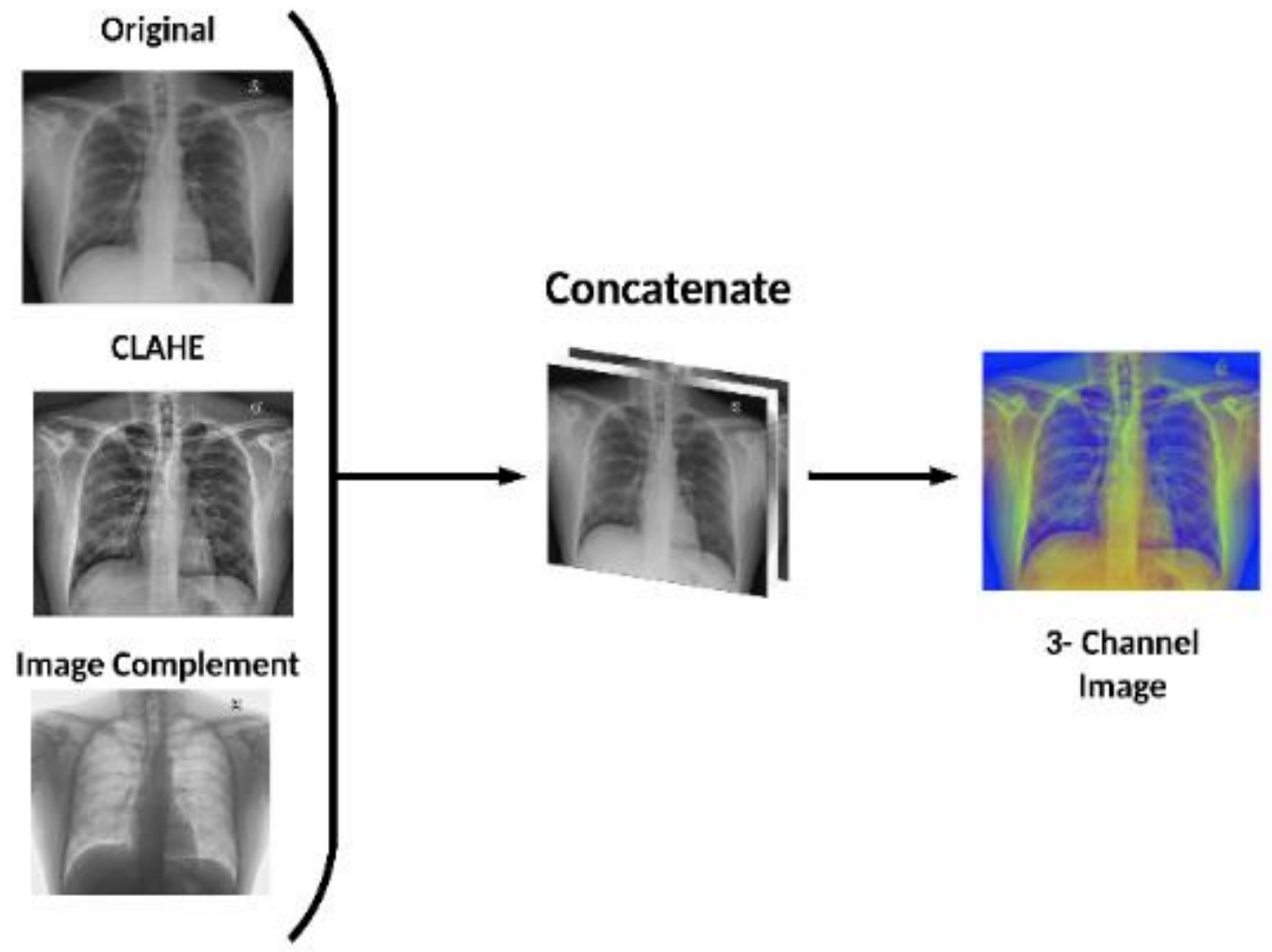

Figure 5. Illustration of 3-channel scheme.

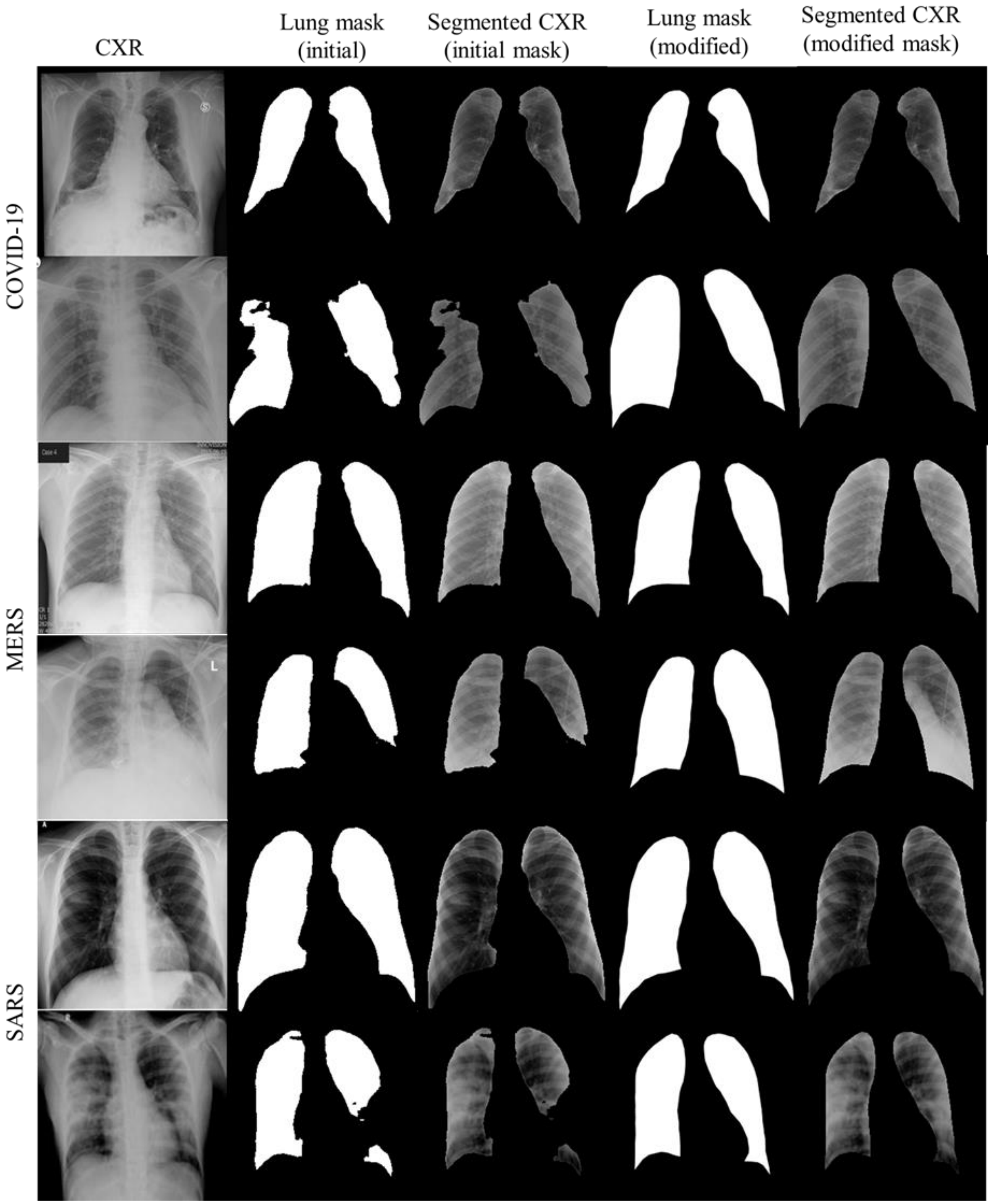

Figure 6. Qualitative evaluation of the U-net model. Original X-ray images (left), lung mask generated by the trained U-net model and corresponding segmented lung, fine-tuned mask by the radiologist and their corresponding lung segment.

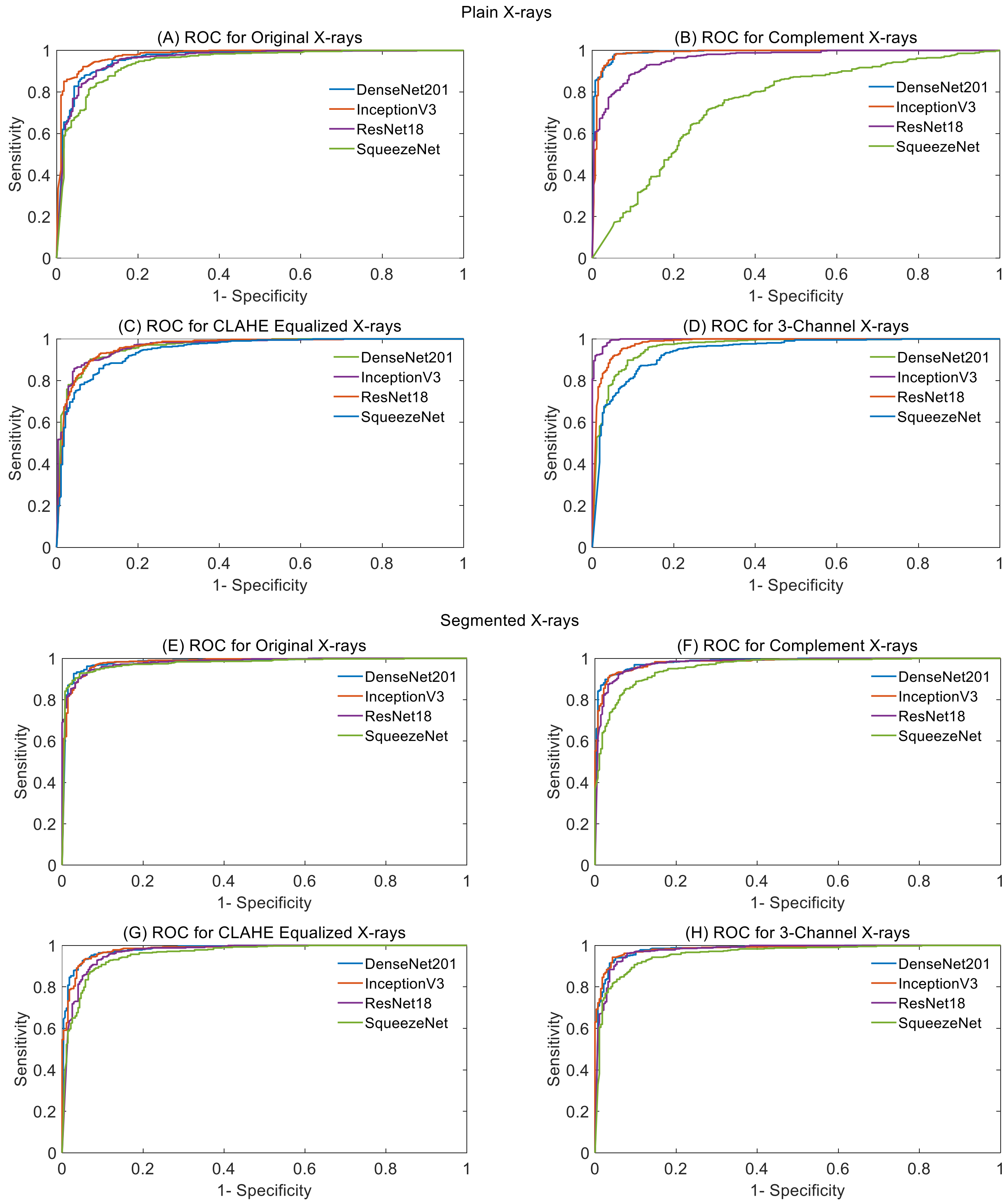

Figure 7. Comparison of the ROC for all folds for four networks using plain X-ray images (A-D) and segmented lung images (E-H): Original images (A/E), Complemented images (B/F), CLAHE images (C/G), and 3-channel images (D/H).

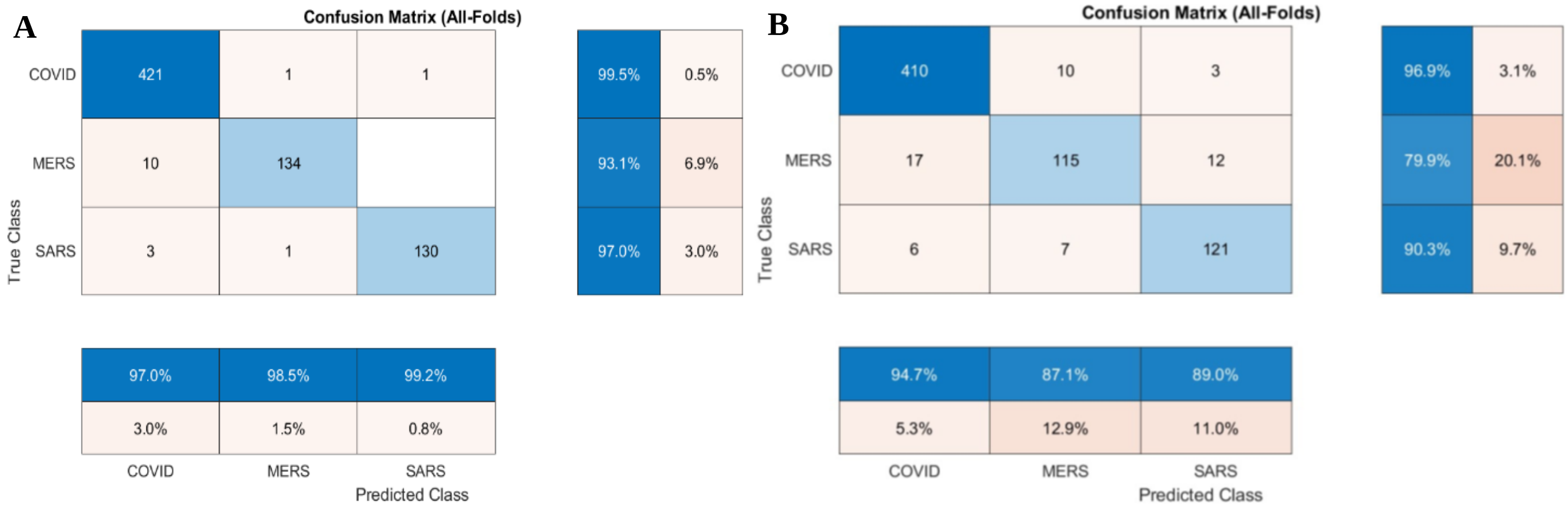

Figure 8. Confusion matrix of all folds for COVID-19, MERS, and SARS for plain (A) and segmented lung (B) X-rays classification using InceptionV3.

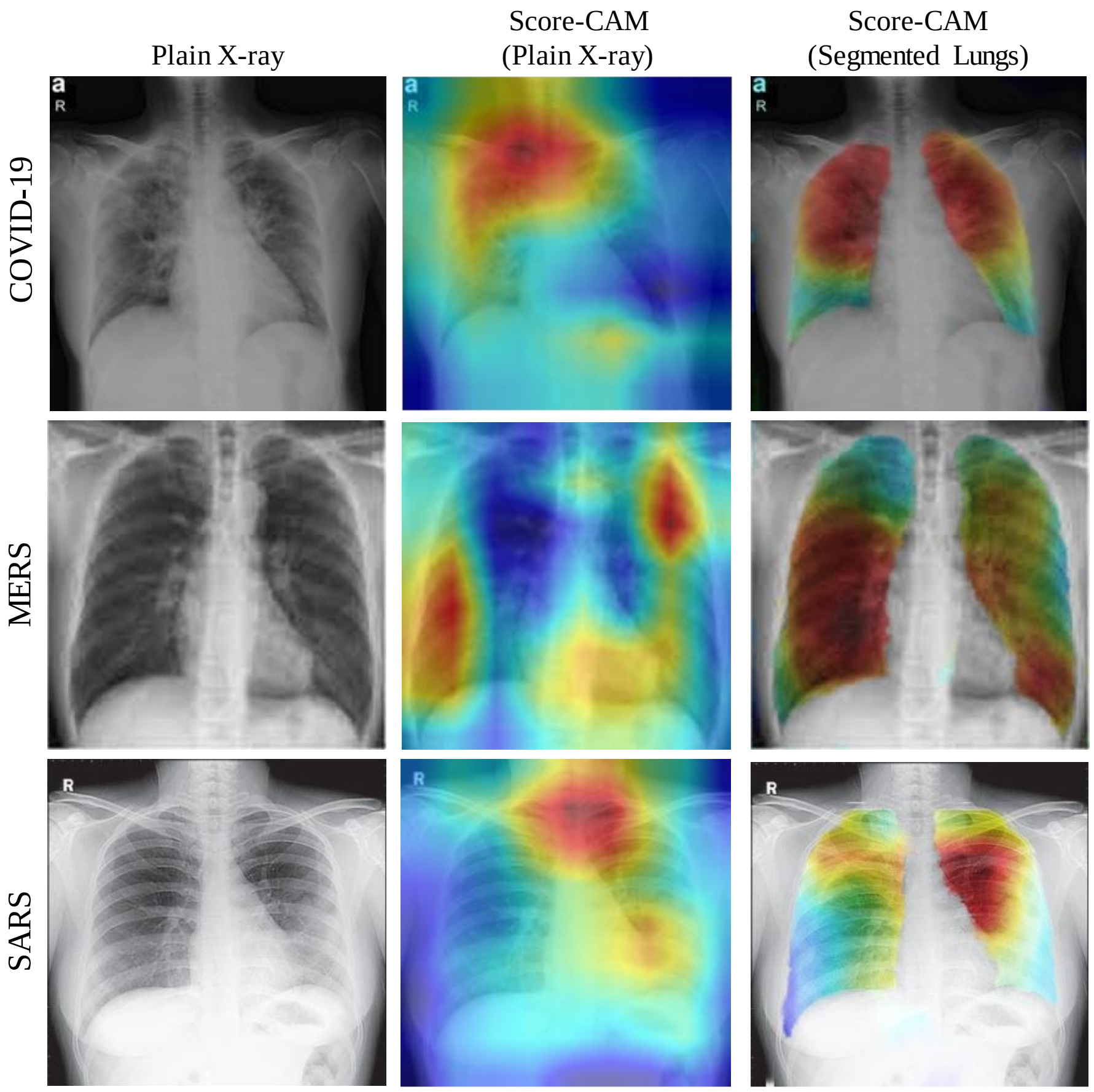

Figure 9. Examples of probabilistic saliency maps for COVID-19, MERS and SARS patients: (A) Plain CXR image, (B) Score-CAM for plain CXR inferred by InceptionV3 network, and (C) Score-CAM for segmented CXR inferred by InceptionV3 network.

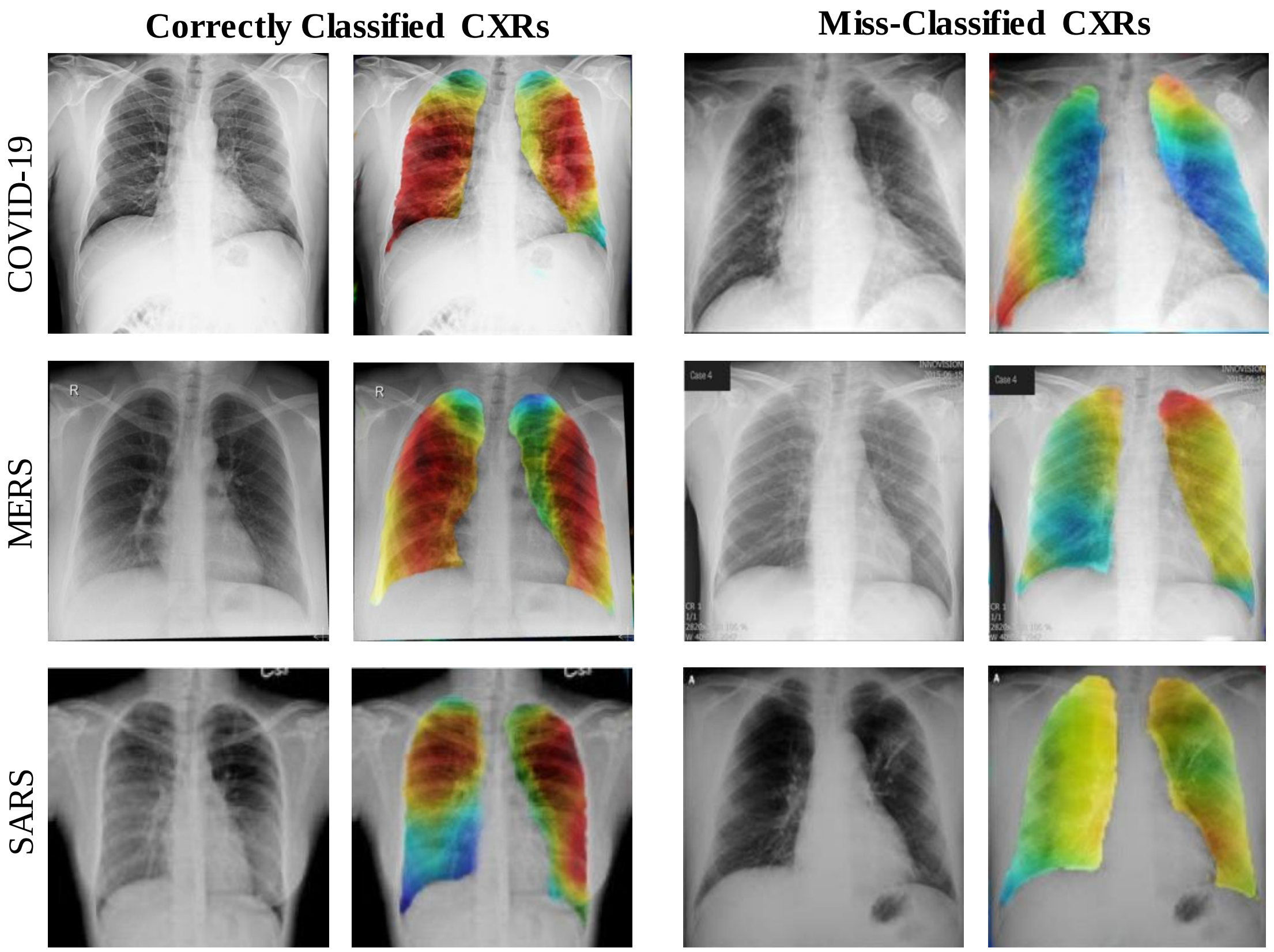

Figure 10. Comparison of the Score-CAM for correctly classified and miss-classified lung CXR images by InceptionV3.